\documentclass[prb,aps,superscriptaddress,reprint]{revtex4-1}

\usepackage{graphicx}
\usepackage[english]{babel}
\usepackage{SIunits}
\usepackage{color}

\begin{document}

\title{Fractional quantum Hall effect in CVD-grown graphene}

\author{M. Schmitz}
\affiliation{JARA-FIT and 2nd Institute of Physics, RWTH Aachen University, 52074 Aachen, Germany, EU}
\affiliation{Peter Gr\"unberg Institute (PGI-9), Forschungszentrum J\"ulich, 52425 J\"ulich, Germany, EU}

\author{T. Ouaj}
\affiliation{JARA-FIT and 2nd Institute of Physics, RWTH Aachen University, 52074 Aachen, Germany, EU}
\affiliation{Peter Gr\"unberg Institute (PGI-9), Forschungszentrum J\"ulich, 52425 J\"ulich, Germany, EU}

\author{Z. Winter}
\affiliation{JARA-FIT and 2nd Institute of Physics, RWTH Aachen University, 52074 Aachen, Germany, EU}
\affiliation{Peter Gr\"unberg Institute (PGI-9), Forschungszentrum J\"ulich, 52425 J\"ulich, Germany, EU}

\author{K. Rubi}
\affiliation{High Field Magnet Laboratory (HFML-EMFL), Radboud University, 6525 ED Nijmegen, The Netherlands, EU}

\author{K.~Watanabe}
\affiliation{Research Center for Functional Materials, National Institute for Materials Science, 1-1 Namiki Tsukuba, Ibaraki 305-0044, Japan}

\author{T. Taniguchi}
\affiliation{International Center for Materials Nanoarchitectonics,National Institute for Materials Science, 1-1 Namiki Tsukuba, Ibaraki 305-0044, Japan}

\author{U. Zeitler}
\affiliation{High Field Magnet Laboratory (HFML-EMFL), Radboud University, 6525 ED Nijmegen, The Netherlands, EU}

\author{B. Beschoten}
\affiliation{JARA-FIT and 2nd Institute of Physics, RWTH Aachen University, 52074 Aachen, Germany, EU}

\author{C.~Stampfer}
\affiliation{JARA-FIT and 2nd Institute of Physics, RWTH Aachen University, 52074 Aachen, Germany, EU}
\affiliation{Peter Gr\"unberg Institute (PGI-9), Forschungszentrum J\"ulich, 52425 J\"ulich, Germany, EU}

\begin{abstract}
We show the emergence of fractional quantum Hall states in graphene grown by chemical deposition (CVD) for magnetic fields from below 3 T to 35 T where the CVD-graphene was dry-transferred.
 Effective composite-fermion filling factors up to $\nu^* = 4$ are visible and higher order composite-fermion states (with four flux quanta attached) start to emerge at the highest fields. Our results show that the quantum mobility of CVD-grown graphene is comparable to that of exfoliated graphene and, more specifically, that the $p/3$ fractional quantum Hall states have energy gaps of up to 
   $30$~K, well comparable to those observed in other silicon-gated devices based on exfoliated graphene.
\end{abstract}


\maketitle

Recent progress in device fabrication of graphene-based heterostructures allows for the observation of an increasing number of composite-fermion (CF) states in graphene~\cite{Bolotin2009Nov,Du2009Oct, Dean2011May, Feldman2012Sep, Feldman2013Aug}. The emergence and the level of manifestation of these states can be regarded as an ultimate proof of device quality in terms of quantum mobility, homogeniety and low residual impurities.
Until present, the observation of such interaction-driven many-body states has been limited to devices based on mechanically exfoliated graphene whereas it has been the subject of discussion whether chemical vapor deposition (CVD) could yield graphene hosting CF states at all.

Dry-transferred CVD-grown graphene has already been demonstrated to be able to compete with the transport properties of highest-quality exfoliated devices with average carrier mobilities exceeding 100,000~cm$^2$/(Vs), mean free paths of up to 28~$\mu$m and well-developed integer quantum Hall states at moderate magnetic fields \cite{Banszerus2015Jul, Banszerus2016Feb, Schmitz2017Jun, Banszerus2019Sep}.
Since these devices are only limited in size by the two-dimensional (2D) materials used for the dry pick-up~\cite{Banszerus2017Jul}, as well as limitations of the copper growth substrates, they might open the door to large area high-quality devices enabling even better understanding of the physics of correlated electron systems resulting in CF states in graphene.
It is therefore of great importance to establish whether the already high carrier mobilities observed in CVD graphene~\cite{Banszerus2015Jul, Banszerus2016Feb,Petrone2012Jun} also translates into sufficiently low residual charge carrier fluctuations, high enough homogenieties and quantum mobilities allowing for the clear emergence of correlation-induced states.

Here we show how to fabricate and characterize  dry-transferred CVD grown graphene heterostructures that show fractional quantum Hall states comparable to exfoliated graphene devices. The activation gaps extracted from these experiments are in good agreement with reported values for silicon-gate based samples and promise even richer CF state spectra when using latest generation of graphitic gating technologies as well as macroscopic devices promised by a CVD graphene based technology \cite{Ribeiro-Palau2019Apr}.

Graphene is grown using low-pressure chemical vapor deposition on the inside of copper foil enclosures~\cite{Li2009Jun,Petrone2012Jun,Chen2013Apr,Hao2013Nov}. The Cu foils are heated up to 1035$^\circ$C within 45 minutes and are further annealed at this temperature for another 15 minutes under a constant flow of 15~sccm H$_2$ gas. This is followed by a 3 hour growth step under a flow of a H$_2$/CH$_4$ mixture (45/5~sccm) at the very same temperature. Afterwards the oven is switched off and the lid of the furnace is opened up for a rapid cool-down keeping the same gas mixture flow as before. This process yields large isolated single-layer graphene flakes with lateral sizes of several hundred micrometers, depending only on the growth time~\cite{Banszerus2015Jul}. After growth, we store the cut and unfolded Cu foils in a water-saturated air atmosphere ($> $99$\%$ relative humidity) at a temperature of 30$^\circ$C for two days. This accelerates the water-mediated oxidation process of the copper surface at the graphene/copper interface~\cite{Luo2017May} and is  of crucial importance to weaken the van der Waals adhesion forces between the graphene and the copper surface \cite{Luo2017May}.

\begin{figure*}[tbh]
\centering
\includegraphics[draft=false,keepaspectratio=true,clip,width=0.75\linewidth]{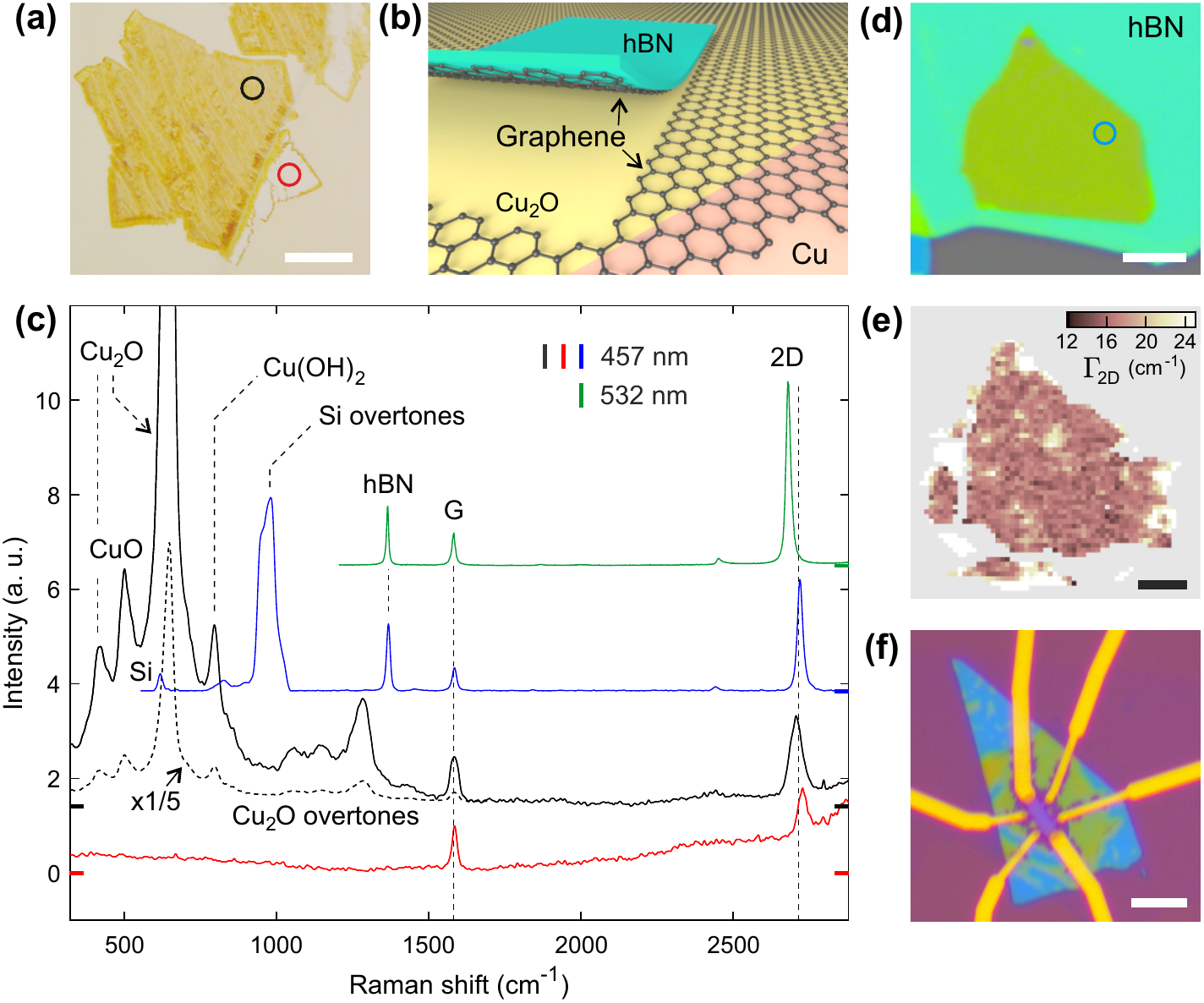}
\caption[Fig01]{\textbf{Device fabrication}  \textbf{(a)} Optical image of CVD-grown graphene (Gr) crystals on a Cu foil after a partially oxidized Cu/Gr interface. Scale bar is 25~$\mu$m. \textbf{(b)} Schematic illustration of the mechanical delamination process.  \textbf{(c)} Raman spectra recorded with two different laser excitation wavelengths (457~nm and 532~nm, see color legend) at the locations highlighted by color matching circles (red and black) on the partially oxidized/non-oxidized copper/graphene areas shown in panel (a) and on the final hBN/CVD-Gr/hBN heterostructure in (see blue and green circle in panel (d)). The spectra are shifted by constant offsets in intensity for comparability (see colored thicks on the right axis) and the black dashed spectrum is a scaled-down version of the black spectrum. \textbf{(d)} Optical image of a finished hBN/CVD-Gr/hBN heterostructure (acale bar is 5~$\mu$m). \textbf{(e)} Raman map showing the 2D line width $\Gamma_{\mathrm{2D}}$ of the graphene layer encapsulated in the heterostructure shown in panel (d). Scale bar is 5~$\mu$m. \textbf{(f)} Optical image of the etched and contacted hBN/CVD-Gr/hBN Hall bar structure (scale bar is 10~$\mu$m).
}
\label{f1}
\end{figure*}

\begin{figure*}[!tbh]
	\centering
		\includegraphics[draft=false,keepaspectratio=true,clip,width=0.68\linewidth]{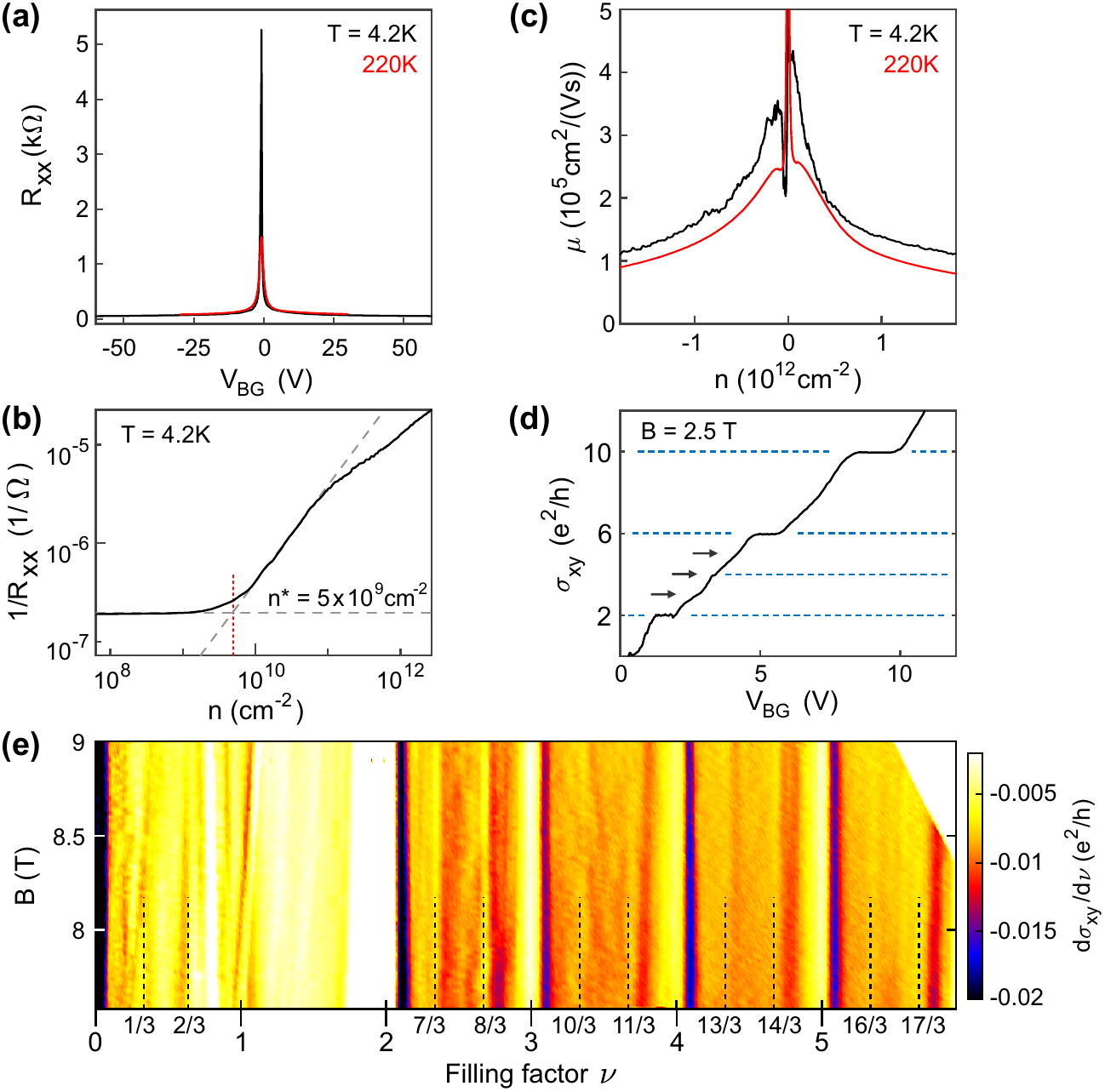}
	\caption[Fig02]{ \textbf{Low temperature magneto-transport measurements}  \textbf{(a)} Four terminal resistance $R_{xx}$ as function of back gate voltage $V_{\mathrm{BG}}$ at a temperature of 220~K (red) and 4.2~K (black). \textbf{(b)} The inverse four terminal resistance $1/R_{xx}$ as function of the charge carrier density $n$. The intercept of two lines following the regimes of constant slope provides an estimate of the residual charge carrier fluctuations $n^* = 5 \times 10^9$~cm$^{-2}$. \textbf{(c)} Charge carrier mobility $\mu$ as function of charge carrier density $n$. \textbf{(d)} Transconductivity $\sigma_{xy}$ as function of back gate voltage $V_{\mathrm{BG}}$ at a perpendicular magnetic field of $B = 2.5$~T ($T=1.8$~K). \textbf{(e)} Landau fan diagram. Differential transconductivity d$\sigma_{xy}/$d$\nu$ as function of filling factor $\nu$ and magnetic field $B$ recorded at $T=1.8$~K.}
	\label{f2}
\end{figure*}

Fig.~1\textbf{(a)} shows an optical image of a CVD-grown graphene crystal with a partially oxidized graphene/copper interface.
The water intercalation and oxidation process can be monitored by confocal Raman spectroscopy when using a laser excitation wavelength of 457~nm (blue), see Fig.~1\textbf{(c)}. We note that a Raman laser excitation wavelength of 532~nm (green) cannot be used to characterize graphene on copper as it creates a broad photoluminesence (PL) emission peak from the underlying Cu, which dominates the emission spectrum \cite{Choi2020Jan}. When using the shorter excitation wavelength, the PL emission from the Cu is shifted to larger wavenumbers thereby allowing to analyze all relevant Raman peaks up to 3000 cm$^{-1}$.
Raman spectra recorded at the oxidized and non-oxidized areas (black and red spectra in Fig.~1\textbf{(c)},  highlighted by color matching circles in Fig.~1\textbf{(a)}) show both the characteristic graphene Raman G- and 2D-peaks at $\omega_G$~=~1582~cm$^{-1}$ and $\omega_{2D}$~=~2720~cm$^{-1}$ \cite{Ferrari2006Oct}. The absence of the defect-related D-peak around 1340~cm$^{-1}$ in the sepctra 
indicates that the graphene is grown defect-free \cite{Ferrari2013Apr} and remains virtually undamaged by the oxidation process. In further comparison of the two spectra, we denote the emergence of a variety of intense peaks visible in the left part of the black spectrum. These peaks can be attributed to Raman-active modes of various copper oxides formed at the graphene/copper interface~\cite{Luo2017May,Galbiati2017Jan}, where the three most prominent peaks are due to cupric oxide CuO (500~cm$^{-1}$), cuprous oxide Cu$_2$O (640~cm$^{-1}$) and copper hydroxide Cu(OH)$_2$ (800~cm$^{-1}$) and the less prominent peaks (422~cm$^{-1}$, 1284~cm$^{-1}$) are overtones of the Cu$_2$O mode.
All these peaks are absent in the red spectrum, which has been taken at a spot where the Cu surface is not yet oxidized.

 \begin{figure*}[!tb]
\centering

\includegraphics[draft=false,keepaspectratio=true,clip,width=0.75\linewidth]{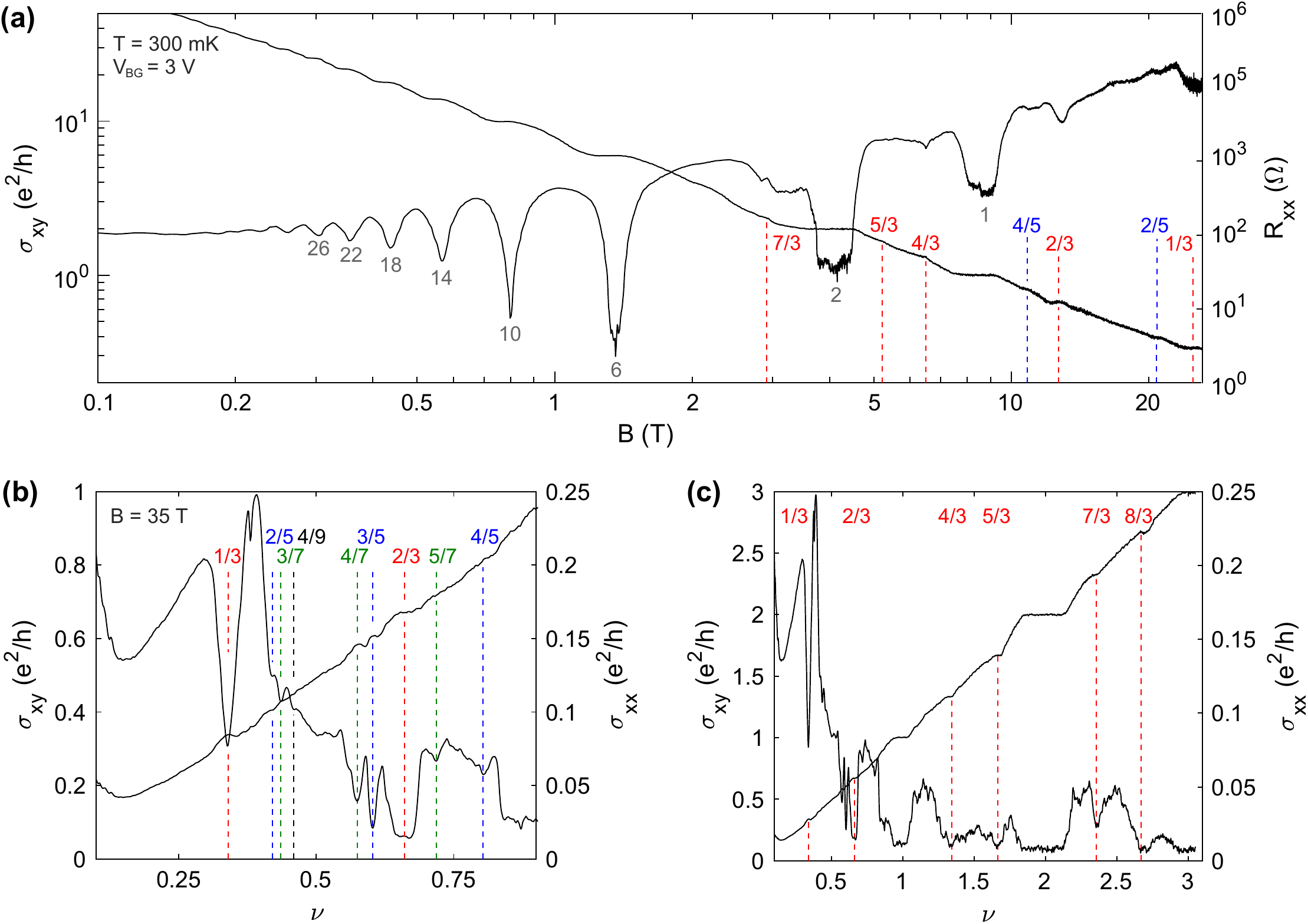}
\caption[Fig03]{ \textbf{High magnetic field measurements}  \textbf{(a)}  Transconductivity $\sigma_{xy}$ and resistance $R_{xx}$ as function of magnetic field $B$. \textbf{(b),(c)} Transconductivity $\sigma_{xy}$ and longitudinal conductivity $\sigma_{xx}$ as function of filling factor $\nu$ at a fixed magnetic field of $B = 35$~T. States with filling factors being integer multiples of $\frac{1}{3}$,$\frac{1}{5}$ and $\frac{1}{7}$ are highlighted in red, blue and green for better visibility.}
\label{f3}
\end{figure*}

As the oxidation process proceeds with time and the oxide layer underneath the graphene grows in thickness, the intensity of these peaks relative to the graphene modes rises. The oxidation process in water saturated air essentially works by the functionalization of the graphene edge by water molecules forming oxygen containing groups like hydroxyl, which in recombination allow the formation of Cu$_2$O \cite{Luo2017May}. This mechanism allows an advanced quantification of the oxidation status of the graphene/copper interface in addition to microscope imaging as the relative intensity of the Cu$_2$O peaks rise with ongoing oxidation. As illustrated in Fig.~1\textbf{(b)}, the graphene crystal can be delaminated from the copper oxide by means of a dry transfer technique using a soft stamp based on hBN~\cite{Banszerus2015Jul} or other 2D materials~\cite{Banszerus2017Jul}.

In this work, the graphene was delaminated from the Cu foil using an exfoliated hexagonal boron nitride (hBN) crystal placed on top of a polymethyl-methacrylate (PMMA) and polyvinyl-alcohol (PVA) stamp, that itself is fixed on a small block of polydimethylsiloxane (PDMS).
Subsequently, the graphene/hBN stack is placed on a second hBN crystal that was previously exfoliated onto a Si$^{++}$/SiO$_2$ substrate. The PDMS block was peeled off and the PVA and PMMA layers were dissolved in hot water (95$^\circ$C) and acetone, respectively. Fig.~1\textbf{(d)} shows an optical image of one of the final hBN/CVD-Gr/hBN heterostructures used in this work \cite{Banszerus2015Jul}.

Again we used confocal Raman spectroscopy to characterize the structural quality of the heterostructures. In Fig.~1\textbf{(c)} we show representative Raman spectra of the sandwich shown in Fig.~1\textbf{(d)} recorded with $\lambda = 457$~nm (blue spectrum) and, for comparison, at the same location, with $\lambda = 532$~nm (green spectrum). The spectra exhibit the characteristic Raman hBN-peak at $\omega_{\mathrm{hBN}} = 1366$~cm$^{-1}$ as well as the Raman G- and 2D-peaks of graphene \cite{Ferrari2006Oct, Graf2007Feb, Ferrari2013Apr, Forster2013Aug}.
The low frequency copper-oxide peaks have disappeared entirely and the D-peak remains absent within the entire heterostructure (note that the appearing Si-peaks are all well understood~\cite{Kallel2014Sep}). This highlights the contamination-free and non-invasive nature of our dry transfer process. A difference of around 20~cm$^{-1}$ in the 2D-peak position $\omega_{2D}$ recorded with the blue and green laser, respectively, can be explained by the energy-dispersive nature of the 2D-peak~\cite{Berciaud2013Aug}.

Since the overall Raman intensity yield from the green laser excitation is much higher than the intensity obtained from the blue laser and to allow a comparison with previous works, we perform further examinations with the green laser only. The 2D-peak exhibits a narrow full-width-at-half-maximum (FWHM) of 17~cm$^{-1}$. This value is comparable to other reported high mobility samples obtained from both exfoliated and CVD-grown graphene \cite{Wang2013Nov, Banszerus2015Jul, Banszerus2016Feb}. The positions of the Raman G- and 2D-peaks around $\omega_G$ = 1582~cm$^{-1}$ and $\omega_{2D}$ = 2680~cm$^{-1}$ point to an overall low doping concentration of the graphene \cite{Lee2012Aug}.
Fig.~1\textbf{(e)} shows a spatial map of the line width of the 2D peak, $\Gamma_{2D}$, of the hBN/CVD-Gr/hBN heterostructures shown in Fig.~1\textbf{(d)}. $\Gamma_{2D}$ is the lowest at regions where the graphene is fully encapsulated in hBN with values 
 around 17~cm$^{-1}$ indicating very small nanometer-scale strain variations over the entire graphene layer~\cite{Couto2014Oct, Neumann2015Sep}.
 
\begin{figure*}[tbh]
\centering

\includegraphics[draft=false,keepaspectratio=true,clip,width=0.65\linewidth]{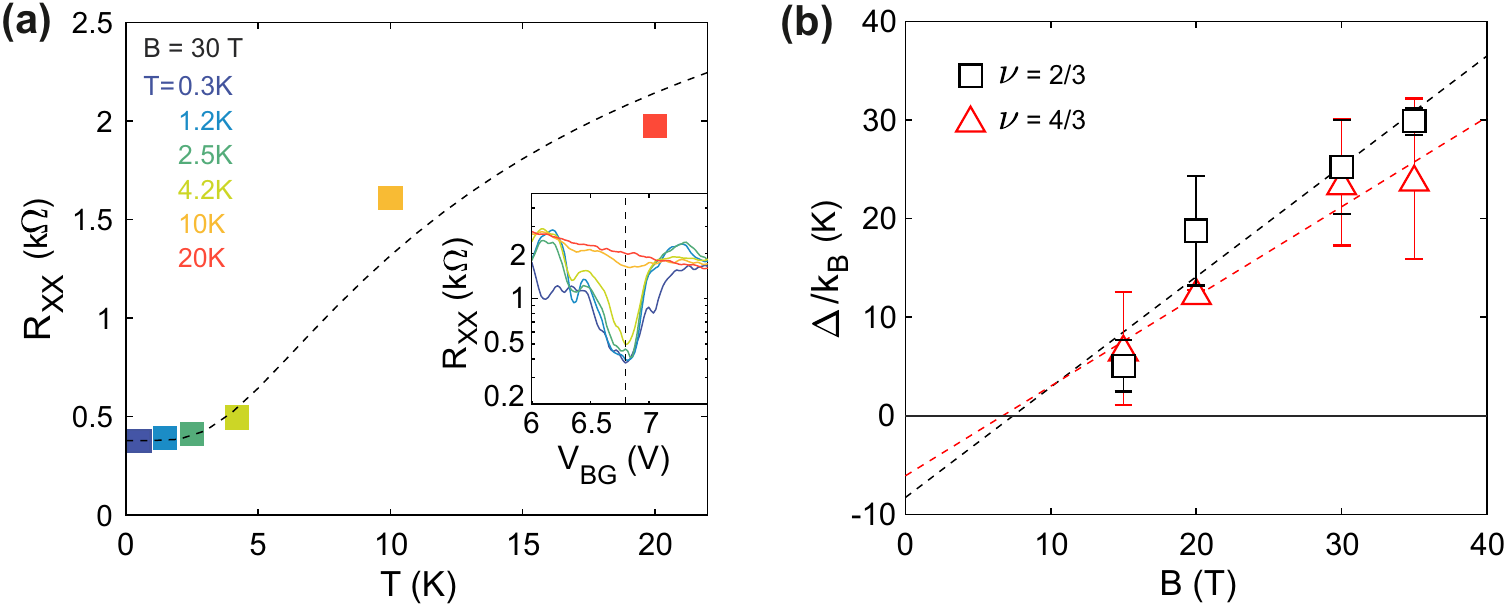}
\caption[Fig04]{\textbf{Activation gap measurements} \textbf{(a)} Bulk resistance $R_{xx}$ of the 4/3 state as function of temperature $T$ together with a fit of a temperature activated model (see text) at a constant magnetic field of 30~T. The inset illustrates the vanishing of the bulk resistance dips when increasing $T$ from 300~mK to 20~K . \textbf{(b)} Extracted activation gaps $\Delta$ as function of magnetic field for the 2/3 and 4/3 fractional quantum Hall states. Dashed lines represent linear fits. }
\label{f4}
\end{figure*}

Areas with the lowest mean Raman 2D line width values are considered for devices. Electron beam lithography as well as SF$_6$-based reactive ion etching (RIE) were employed to fabricate Hall-bar-shaped structures with a channel of 2.3~$\mu$m in width and 4~$\mu$m  in length. The Hall-bar structures were then covered by an additional hBN crystal to protect the devices from additional contaminations and quality degradation over time. A second RIE step and metal evaporation were performed to create one dimensional Cr/Au (5~nm/95~nm) contacts. An optical image of the final device is shown in Fig.~1\textbf{(f)}. For electrostatic gating, we use the highly p-doped silicon at the back side of the substrate that is covered by a 300~nm thick silicon oxide layer, where the contacted sandwich structure is resting on.
Taking the thickness of the bottom hBN crystal into account this leads to a back gate (BG) capacitance per area $A$ of $C_{\mathrm{BG}}/A = \varepsilon_0 \varepsilon_r/e(d_\mathrm{Si}+d_\mathrm{hBN}) \approx 6.4 \times 10^{10}$~cm$^{-2}$/V, where $\varepsilon_0$ is the dielectric constant, $\varepsilon_r \approx 4$ is the relative permittivity of hBN and SiO$_2$, $e$ is the electron charge and $d_\mathrm{Si}$~=~300~nm and $d_\mathrm{hBN}$~=~45~nm are the thicknesses of the hBN crystal and the SiO$_2$ layer.

Transport measurements were performed in a pumped
$^4$He cryostat system with a base temperature of 1.8~K ($B$-fields up to 9~T) and in a
$^3$He cryostat system with a base temperature of 300~mK ($B$-fields up to 35~T).
We used standard lock-in techniques to measure the four-terminal resistance
as well as the Hall voltage for extracting the Hall conductivity.

When cooled down to 4.2~K the device showed a low overall doping ($n = 5\times10^{9}$~cm$^{-2}$,  with the charge-neutrality point (CNP) situated close to zero gate voltage, $V_{\mathrm{CNP}}=~-0.8$~V) and low residual charge carrier inhomogeneities of $n^* = 5\times 10^{9}$~cm$^{-2}$, see Figs.~2\textbf{(a)} and \textbf{(b)}. These findings are in good agreement with spatially resolved confocal Raman measurements on the heterostructure as well as with earlier works on dry-transferred CVD graphene~\cite{Banszerus2015Jul,Banszerus2016Feb}.

Fig.~2\textbf{(c)} shows the charge carrier mobility $\mu$ as function of charge carrier density $n$ for two different temperatures extracted using the Drude formula $\sigma = ne\mu$, where $\sigma$ is the electrical conductivity and $e$ is the electron charge. Close to the CNP, the mobility at 4.2 K increases to values around 350,000 and 400,000~m$^2$/(Vs) and around 250,000~cm$^2$/(Vs) at 220~K. In agreement with earlier work \cite{Ribeiro-Palau2019Apr}, we observe a slight discrepancy in quality for the two carrier types and will therefore only focus on the electron governed regime for the remainder of this work.

After having demonstrated the high electronic quality of the device, we have carried out magnetotransport measurements with the magnetic field applied perpendicular to the graphene layer. Fig.~2\textbf{(d)} shows the Hall conductivity $\sigma_{xy}$ as function of back gate voltage $V_{\mathrm{BG}}$ at a fixed magnetic field of 2.5~T. Even at this low field, the onset of degeneracy lifting into integer quantum-Hall states is well visible.
This lifting becomes more clearly at higher magnetic fields as shown by
the Landau fan diagram in Fig.~2\textbf{(e)} which shows a false-color plot of the differential Hall conductivity $d\sigma_{xy}/d\nu$ as function of the filling factor $\nu$ and the magnetic field $B$. The former is given by $\nu = (V_{\mathrm{BG}}-V_{\mathrm{CNP}})\alpha h/(eB)$ with $V_{\mathrm{BG}}$ being the back gate voltage and $h$ the Planck constant.
The lever arm of the back gate $\alpha = 6.34 \times10^{10}$~cm$^{-2}$/V was extracted from quantum Hall measurements and is in good agreement with the value obtained from the parallel plate capacitor model ($C_{\mathrm{BG}}/A$; see above).
The Hall conductivity  displays well-developed integer quantum Hall plateaus and furthermore reveals clear indications of states at fractional filling factors $p/3$ even below 8~T, as highlighted by the dashed lines.

Fractional quantum Hall (FQH) states can be attributed with a composite fermion filling factor $\nu^\ast$, that can be related to the integer quantum Hall filling factor $\nu$ by $\nu = \nu^\ast/(2p\nu^\ast \pm 1)$, where $2p$ is the vorticity or number of flux quanta attached to one composite fermion. In order to study these fractional states in more detail, the device was further characterized at magnetic fields up to 35~T. 
Fig.~3\textbf{(a)} shows the four terminal resistance $R_{xx}$ (right axis) as well as the Hall conductivity $\sigma_{xy}$ (left axis) as a function of $B$ (logarithmic scale) at a fixed back gate voltage of $V_\mathrm{BG}=3$~V and a temperature of 300~mK.
First signatures of the 7/3~state already appear below 3~T and more clearly pronounced FQH states start to develop at higher magnetic fields.

Figs.~3\textbf{(b)} and 3\textbf{(c)} show the Hall conductivity $\sigma_{xy}$ (left axis)and the corresponding longitudinal conductivity $\sigma_{xx}$ (right axis) as a function of filling factor $\nu$ at $B=35$~T. Well developed plateaus at all $\nu = p/3$ fractional filling factors are visible. Moreover, we can identify all 2-flux CF states ($2p =2$), up to $\nu^{\ast} = 4$ ordered symmetrically around half filling in $\sigma_{xy}$ as well as in $\sigma_{xx}$. Even more, we denote plateaus at $\nu = $5/7 and 4/5 which can be identified as 4-flux CF states ($2p = 4$) centered around $\nu=3/4$.

These results indicate an outstanding material quality considering the fact, that a rather simple Si$^{++}$/SiO$_2$ gate/gate oxide was used instead of the state-of-the-art graphitic gating approaches, which allow for precise tuning of the sample edge \cite{Overweg2018Jan, Banszerus2018Aug, Zibrov2018Jul, Kim2018Nov} or omit the edge completely when using Corbino geometries or electrostatically defined channels \cite{Yan2010Nov, Zeng2019Apr, Chen2019Jan}.

In order to further quantify the robustness of the observed FQH states, we performed temperature dependent activation gap measurements for the most pronounced fractional filling factors $\nu=2/3$ and $\nu=4/3$. A set of back-gate traces was recorded at constant magnetic fields for a range of temperature steps. An Arrhenius-like temperature activated model is used to fit the data where the minimum of $R_{xx}$ at a given filling factor $\nu$ is assumed to follow $R^\nu_{xx,min} \propto$ $R\exp(-\Delta_{\nu}/2k_BT)$, where $k_B$ is the Boltzmann constant and $\Delta_{\nu}$ is the activation gap. Fig.~4\textbf{(a)} shows a fit of the $R_{xx}$ minimum at 4/3 filling as a function of temperature between 300~mK and 20~K at $B=30$~T; the corresponding resistance traces are shown in the inset.  The extracted results for $\Delta_{2/3}/k_B$ and $\Delta_{4/3}/k_B$ as function of the magnetic field are shown in Fig.~4\textbf{(b)}. The activation gaps increase roughly linearly with magnetic field reaching a value of 29~K and 24~K, respectively, at $B=35$~T. This behaviour is in good quantitative  agreement with other studies on FQH states in graphene, where values in the range of 30~K to 60~K were reported for $\Delta_{2/3}/k_B$ \cite{Ribeiro-Palau2019Apr, Zeng2019Apr} and 16~K for $\Delta_{4/3}/k_B$~\cite{Dean2011May}. Assuming the $B$-field dependence being purely given by the CF Zeeman energy (reduced by Landau level broadening) following $\Delta_{\nu} = \frac{1}{2}\mu_BgB+\Gamma_{\nu}$, where $\mu_B$ is the Bohr magneton and $g$ is the Land\'e $g$ factor, we can interpret the intercepts $\Gamma_{\nu}$ as a measure of the disorder-induced Landau-level broadening. In our device we extract $\Gamma$ values between 6 and 8~K, which is in agreement with values of around~10~K on mechanically exfoliated graphene reported in other studies~\cite{Dean2011May, Ribeiro-Palau2019Apr, Zeng2019Apr}. The extracted values of the $g$ factor, ranging between 2.7 and 3.4 are roughly a factor 2 smaller compared to values 
reported in Refs.~\cite{Ribeiro-Palau2019Apr, Zeng2019Apr}. Since the $g$ factor is known to strongly depend on the electron-electron interaction renormalization~\cite{Schulze-Wischeler2004Apr} this value may crucially depend on the electrostatic environment substantially influenced by the different gating technologies.

In summary, we have demonstrated the presence of fractional quantum Hall states in CVD-grown graphene underlining the high electronic quality of this material when assembled into heterostructures by a dry transfer process. Neither CVD growth on Cu substrates, nor the oxidation of the Cu surface via water intercalation nor the dry transfer process induce enough disorder to quench the interaction-driven formation of composite fermions. Even though using a conventional silicon gate to tune the carrier density, CF states emerge even at moderate magnetic fields below 3~T. The full spectrum of 2-flux states up to $\nu^{\ast}$=~4 could be identified at $B = 35$~T as well as several 4-flux states underlining the high quality of the material. The results are unambiguously proving that the high quality of CVD-grown graphene equals that of mechanically exfoliated graphene flakes. Therefore, a new generation of large-area ultra high quality devices might be achieved using scaleably grown 2D materials to dry transfer CVD graphene and open the door to a deeper understanding of the interaction effects driving the emergence of CF states in graphene.

{\it Note.} During preparation of this manuscript, we became aware of a very recent preprint showing indications of the $-1/3$ state in transferred CVD graphene~\cite{Pezzini2020May}.

{\bf Acknowledgements:}

This project has received funding from the European Union's Horizon 2020 research and innovation programme under grant agreement No. 881603 (Graphene Flagship) and from the European Research Council (ERC) under grant agreement No. 820254, the Deutsche Forschungsgemeinschaft (DFG, German Research Foundation) under Germany's Excellence Strategy - Cluster of Excellence Matter and Light for Quantum Computing (ML4Q) EXC 2004/1 - 390534769,
through DFG (BE 2441/9-1 and STA 1146/11-1), and by the Helmholtz Nano Facility~\cite{Albrecht2017May} at the Forschungszentrum J\"ulich.
K.W. and T.T. acknowledge support from the Elemental Strategy Initiative conducted by the MEXT, Japan, Grant Number JPMXP0112101001,  JSPS KAKENHI Grant Number JP20H00354 and the CREST(JPMJCR15F3), JST.
Part of this work was supported by HFML-RU/NWO-I, member of the European Magnetic Field Laboratory (EMFL).

\end{document}